\documentclass[12pt]{iopart}

\newcommand{\bit}{\begin{itemize}}
\newcommand{\eit}{\end{itemize}}
\newcommand{\beq}{\begin{equation}}
\newcommand{\eeq}{\end{equation}}
\newcommand{\ben}{\begin{enumerate}}
\newcommand{\een}{\end{enumerate}}

\renewcommand{\theenumi}{\arabic{enumi}}

\usepackage{xcolor}
\usepackage{amssymb}
\usepackage{amsmath}
\usepackage{siunitx}
\usepackage{outlines}
\usepackage{graphicx}

\usepackage[colorlinks=true, linkcolor = black, citecolor = black, urlcolor = blue!60!black]{hyperref}

\begin{document}

\title[Qualitative observations in university physics laboratories\ldots]{Qualitative observations in university physics laboratories: an example from classical mechanics}

\author{K. Dunnett and M. H. Magnusson}
\address{Department of Physics, Lund University, Box 118, 221 00 Lund, Sweden}
\ead{martin.magnusson@ftf.lth.se}
\vspace{10pt}
\begin{indented}
\item[] \today
\end{indented}

\maketitle

\begin{abstract}
One of the key skills of a researcher is noticing what's going on. Both in the experiment one's performing and in one's data: is there something interesting, reason to doubt one's data or suspect that one's theoretical description is insufficient? Many experiments developed for undergraduate teaching still focus on quantitative evaluation. Here we take an alternative approach where careful observation identifies the interesting qualitative behaviour of a ball dropped with a water bottle balanced on top of it, but where numerical agreement with a simple theoretical model is impossible. Thus `success' occurs when students are satisfied with their efforts and the development of their experimental process. Laboratory note keeping can also be introduced in a meaningful, non-formulaic way since students are making independent observations and method changes. We describe pedagogical and didactic considerations for the implementation of the experiment in a classroom, including variations and extensions, and give examples of experimental outcomes. We suggest that considering qualitative behaviour may be a fruitful strategy for identifying experiments that are both amenable to student autonomy and embedding skills such as laboratory note keeping in a flexible and genuine way.

\end{abstract}

\section{Introduction}

Practical work is considered a cornerstone of work in the natural sciences, and therefore a necessary part of scientific training \cite{SciEd.88.28, RevEdRes.52.201}. However, practical work is often expensive in terms of staffing, space and equipment, and many activities, especially at the introductory level are very rigidly instructed \cite{ChemEdResPract.8.212, JColSciTeach.38.52}. This format does not mimic investigative science \cite{JEngEd.100.540} and repeated calls have been made for more `open' or genuine practical work \cite{SciEd.88.28, RevEdRes.52.201}. More recently, impacts of highly instructed practical work, that may be deemed undesirable from the perspective of training scientific attitudes, have been reported \cite{PRPER.18.010128, PRPER.16.010113, PNAS.112.11199, PRPER.17.020112}. Investigative \cite{PRPER.12.020132, PRX.10.011029, IntJScholTandL.3.2.16} and skills-based \cite{PRPER.18.010128} practical work (more generally, activities allowing student agency \cite{PRPER.16.010109, PRPER.17.020128}) have been demonstrated to support students in developing scientific thinking skills, while highly instructed practical work does not well support students developing conceptual understanding \cite{PRPER.13.010129}. Despite this, practical activities with the goal of reinforcing or supporting conceptual understanding through demonstrating phenomena still abound \cite{PRPER.16.020162, PRPER.18.020129}. Moreover, when faced with practical work where numerical results will deviate from the idealised theory presented, some students will strategically avoid or otherwise ignore the discrepancies \cite{PRPER.16.010113, PRPER.17.020112, EJP.40.015702}.

In the existing rubric for characterising practical work as `confirmation', `structured', `guided', `open' or `authentic' inquiry, moving from `confirmation' towards `authentic inquiry' consists of removing instructor predetermination from the end of an activity \cite{JColSciTeach.38.52}. However, investigation can also be supported within activities where there is an expected numerical outcome or known features of interest \cite{TPT.53.352, ProcPERC.2020.539, EJP.45.055201}. In these cases, students can have significant freedom in how the data is collected, but the discussion is to a degree guided by staff \cite{TPT.53.352, ProcPERC.2020.539}. The vast majority of examples to date still focus on measurement taking, with one of the legitimate strategies for students who are determined that the data must agree with the theoretical description being to (artificially) decrease the trustworthiness of their data \cite{PRPER.16.010113, PRPER.17.020112}.

One objection sometimes made to inquiry practical work is that students do not engage fully with opportunities designed into tasks \cite{PRPER.16.010113, PRPER.17.020112}, especially in grade-focussed systems \cite{DeFeo2021, EJP.40.015702}, which therefore makes such activities pointless to run. Another objection sometimes made is that inquiry practical work is associated with unreasonable cognitive load for students \cite{EdPsy.41.75}, although this can be addressed through careful activity design and appropriate scaffolding \cite{EdPsych.42.99, PRPER.17.010131}. In particular, given that practical work doesn't contribute significantly to students developing conceptual understanding \cite{PRPER.13.010129}, the focus of practical work should consider other learning that can occur in a laboratory context (e.g. \cite{TPT.57.296}). Specifically, a careful consideration of the (possibly unstated) intended learning outcomes \cite{Biggs-and-Tang} of a practical task can be useful. In the following we present a simple experiment (low prior knowledge requirement) suitable for introductory university physics practical work where the focus is on qualitative behaviour and observation (skills focus and no theory to agree with) as an example of an experiment that can address some of the reported challenges with inquiry practical work, while providing students with ample opportunities to practice scientific critical thinking.  

Writing is an important academic skill that students are expected to develop during university studies, although many may not appreciate how pervasive this is \cite{AffectRes.2022.35}. Physics is stereotypically associated with more mathematical calculations and experimental data collection than extensive amounts of writing. Note keeping during experimental and other investigative work is an essential skill that many students do not receive any (useful) structured training in \cite{PRPER.12.020129}. However, in pre-university studies, students may have encountered writing in connection to physics predominantly in the form of (partial) formal reports on experimental work \cite{AlfonsMagnus-2023}. It is almost self-evident that if one has nothing written \emph{down}, one has very little to write \emph{up}. However, when asked to keep notes in lab our students often appeared uncertain of what they should write. With experience of systems where laboratory note[book] keeping is emphasised and introduced through formal guides resulting in many students anxiously spending large amounts of time confirming that they are recording the `correct' information in the right way, a less formulaic way to introduce note keeping was desired. Early in the development of this lab exercise, we realised that its observational and experimental nature could provide an authentic and meaningful introduction to lab note keeping (note, no books necessary). This was embedded into the subsequent development. The implementation described here therefore incorporates three distinct areas of experimental skills development: observation, method development, and note keeping.

\section{A ball and a bottle}

\subsection{The basic idea of the experiment}

The experiment is based on the observation that when an empty \SI{500}{ml} plastic bottle is placed atop a leather football and both are released together and bounce, the bottle flies very high. As described in more detail in section \ref{theorysec}, this situation exemplifies an light object bouncing off a [moving] heavy target, a scenario commonly discussed in physics and mechanics textbooks \cite[Chapter 9]{WalkerResnick-book}. Adding water to the bottle turns this observation into an investigation of the dependence of the bounce heights on the mass of the `lighter' object. In this paper, \textit{bounce height} refers to \textit{the maximum height the object reaches after the first bounce off the ground}. The qualitative result is that the bottle's bounce height decreases for larger masses. More interestingly, the ball's bounce height reaches a minimum (very close to zero) when the bottle is part full of water \cite[Example 9:69]{WalkerResnick-book}, and with more water in the bottle, the ball's bounce height increases. Another point of interest occurs when the ball and the bottle have approximately the same mass and they bounce as a single object (they do split soon after). 

The reason for conducting a qualitative experiment rather than a quantitative one is that trying to take precise measurements might distract students from \textit{observing} what happens. The interesting phenomenon is that there \emph{is} a minimum in the ball's bounce height, not precisely \textit{how high} the ball bounces. In a practical setting, this experiment is very difficult to reproduce precisely, making quantitative data not very useful (cf. Fig. \ref{fig:quant-plot}); requiring students to take numerical measurements is likely to lead to more frustration than understanding.

\subsection{The experiment in the classroom
\label{sec-classroom}}

This experiment works best with several groups of 3--4 students (not less than 3 students per group because of practical issues; more than 4 students per group is also not ideal since there would not be enough for people to do). The total number of students that can be practically facilitated in a session will depend on the space available and `teacher' experience. The experiment works best with an area of at least 4 $\times$ \SI{4}{m} per group, and sufficient ceiling height -- the space and height required can be reduced slightly by using low release heights and making sure that all students are alert to balls and bottles from neighbours as well as their own groups. Each group should also have access to a whiteboard or similar large writing surface. At the beginning of the session, students are shown a demonstration using an empty bottle (and non-optimised, single person release) to provide them with an idea of the basic behaviour, and the extreme range of the bottle's flight. They are then set to do the investigation with minimal instruction.

The approximate time allocated for the experimentation is \SI{75}{min} (plus \SI{15}{min} introduction and 15--\SI{30}{min} for the concluding discussion). During the experimentation, teaching assistants circulate, offering hints when needed (mostly on observed methodological difficulties, and a prompt to repeat or refine the experiment if their measurements do not include the known minimum in the ball's bounce height), and reminding students to keep notes (observations, method changes etc.). The expectation is that all groups will take measurements that reveal the minimum in the ball's bounce. The instructions are deliberately minimal, since `unsuccessful' outcomes are valuable for group discussions, and there are other valid variations on the experiment that should be encouraged if a group starts doing them. Getting the bottle to the same mass as the ball requires careful measurement and somewhat counter to the observational focus of the experiment, however, one can ask students how they could be sure of $a=1$ without measurement given that there are several sets of equipment. It will take a few attempts to release two balls sufficiently straight for them to bounce together, but it is doable, and the single object-like behaviour is clear.

Following the experimental period, students' results, observations and method modifications, are discussed with all students in the session. Some discussion of the relevant physics concepts and processes may also occur, but a detailed explanation of the physics is not the focus in the single-session implementation described here; prompts to reflect upon the physics should \textit{not} be included in the instruction, but can be provided once all the data is collected if a group has lots of time before the plenary discussion.  For more details on the instructions and teacher prompts, see the appendix \ref{SM-classroom}.

\subsection{Theoretical description of the ball and bottle bounce \label{theorysec}}

\begin{figure}[htb!]
    \includegraphics[width = 10cm]{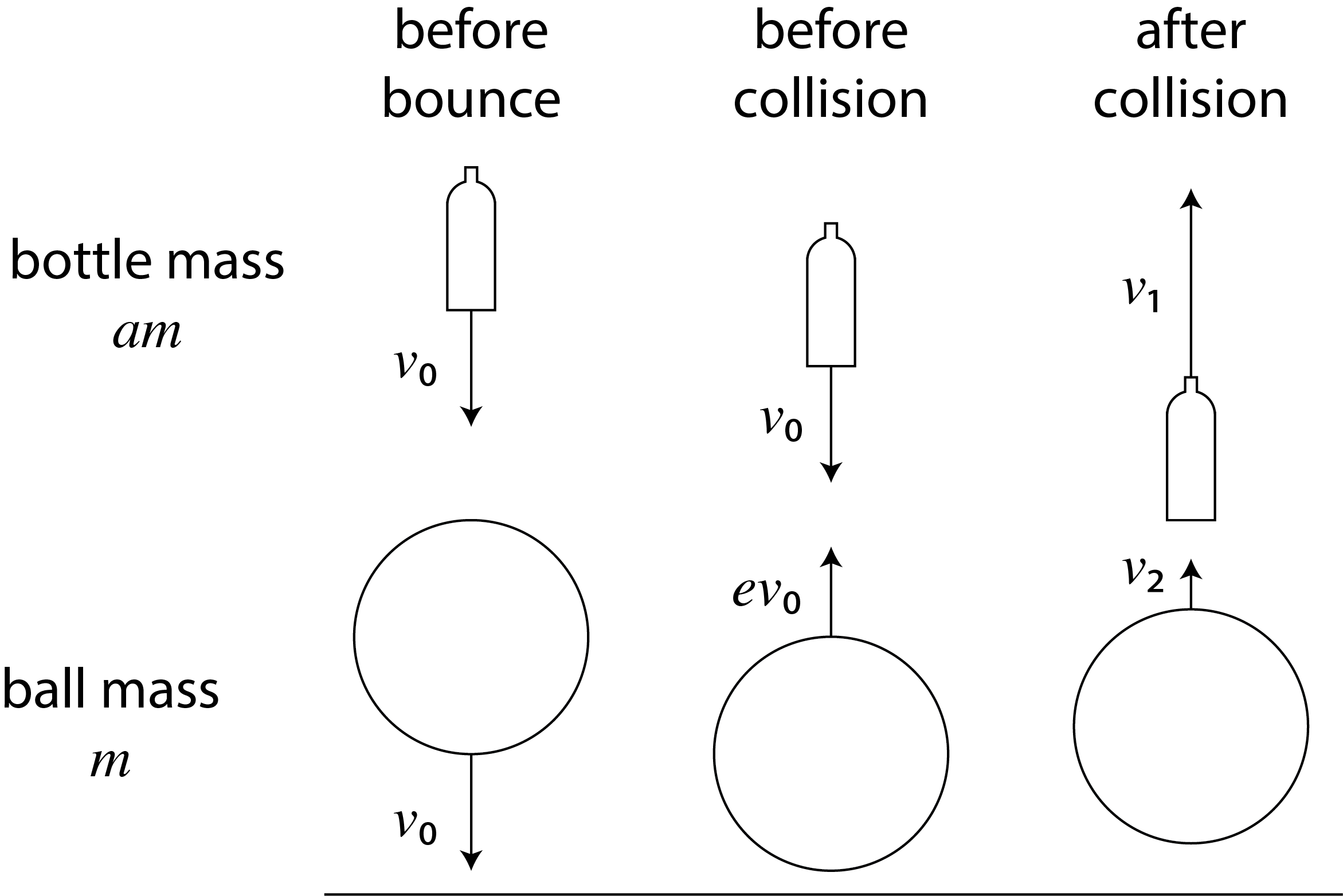}
    \centering
    \caption{The geometry of the ball and bottle bounce experiment, illustrating that the process is a double collision. The left is the situation just before the ball hits the ground. The middle is after the ball has bounced (but before the bottle hits the ball) and the right is after this second collision. The masses and velocities are those used in the full theoretical derivation in the supplemental material \ref{SM-theory}.
    \label{fig:BB-illustration}
    }
\end{figure}

In this paper, the situation is modelled as two separate collisions, rather than as one object which fragments upon impact \cite[Chapter 9]{WalkerResnick-book}. Not only is this model simpler to calculate, it reflects that both masses have some elasticity and will deform upon impact. This means that the top object will continue moving downwards, compressing the the bottom object, even after the latter has reversed direction after the bounce, leading to a time delay. This is true even in the case of nearly incompressible objects, where the time delay would be due to the speed of sound. In short, in the model, the bottle and ball are released at the same time, but one assumes a small gap between them. When the ball hits the ground, it reverses direction and then hits the bottle which is still falling downwards. This geometry is illustrated in Fig. \ref{fig:BB-illustration}; the gaps between the ball and bottle are exaggerated. Assuming that the collision is central and instantaneous, the derivation is straightforward - see \ref{SM-theory}.

For $a=m_\mathrm{bottle}/m_\mathrm{ball}$, with $e$ the coefficient of restitution (elasticity) for each bounce or collision, and $h_0$ the initial release height of the ball and bottle, the resulting bounce heights are:

\beq
    h_\mathrm{bottle}= \left( \frac{e^2+2e-a}{1 + a} \right)^2 h_0, \quad h_\mathrm{ball} = \left( \frac{e - a(1 + e + e^2)}{1 + a} \right)^2 h_0
    \label{inelastic-result}
\eeq

In the case where $a > e/(1 + e + e^2)$ (and especially for $a \gtrsim 1$), the derivation is not fully valid due to multiple collisions, but the qualitative result and the existence of a minimum in the ball bounce height are not affected by this simplification. 

In the fully elastic case ($e=1$), the bounce heights are:

\beq
    h_\mathrm{bottle}= \left( \frac{3-a}{1 + a} \right)^2 h_0, \quad h_\mathrm{ball} = \left( \frac{1 - 3a}{1 + a} \right)^2 h_0
    \label{elastic-result}
\eeq

\begin{figure}[htb!]
\centering{
    \includegraphics[width = 0.75\textwidth]{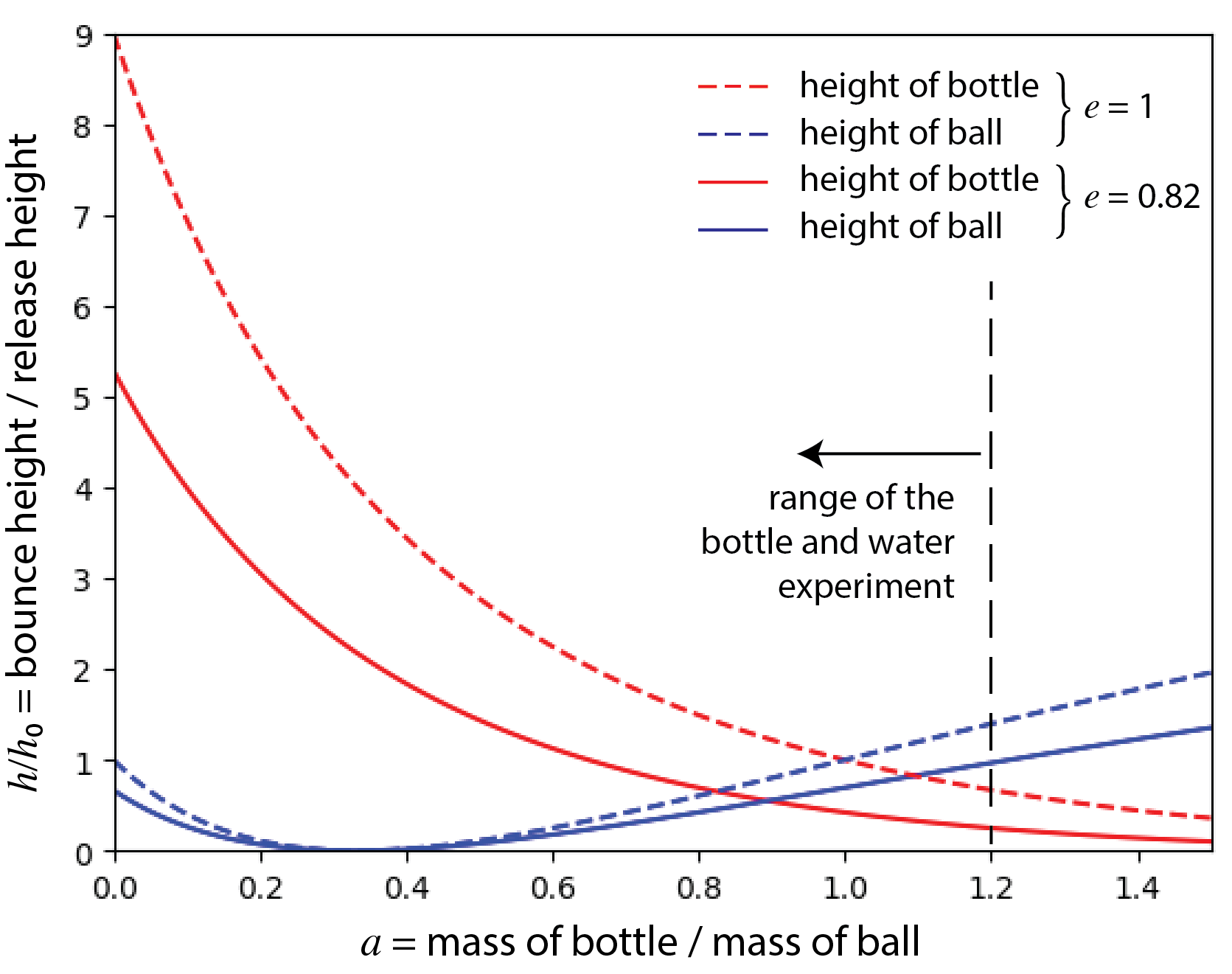}}
    \caption{Bottle and ball bounce heights as a function of $a$, for $e=1$ (Eq. \eqref{elastic-result}) and $e = 0.82$ (Eq. \eqref{inelastic-result}). The release height is $h_0$. The football weighs \SI{425}{g} and an empty bottle \SI{26}{g}, so for a bottle with \SI{500}{ml} of water, $m_{\textrm{bottle}} \approx 1.2 \ m_{\textrm{ball}}$.
    \label{fig:Diagram-TheoryPlot}
    }
\end{figure}

Thus, $ h_\mathrm{bottle} = 9 h_0 $ for an empty bottle ($a \approx 0$); $h_\mathrm{ball}=0$, \textit{i.e.} the ball is fully damped, for $a = {1}/{3}$. For $a = (3-e)/(1+e)$, the ball and bottle have the same bounce height; when $a$ is larger, the ball would bounce higher than the bottle, but in reality this leads to a second collision. Figure \ref{fig:Diagram-TheoryPlot} shows the bottle and ball bounce heights as a function of $a$ and for two different values of $e$. The value $e = 0.82$ is based on the quantitative experiment plotted in Fig. \ref{fig:quant-plot}.

\subsection{Example data}
\begin{figure}[htb!]
    \includegraphics[width = \textwidth]{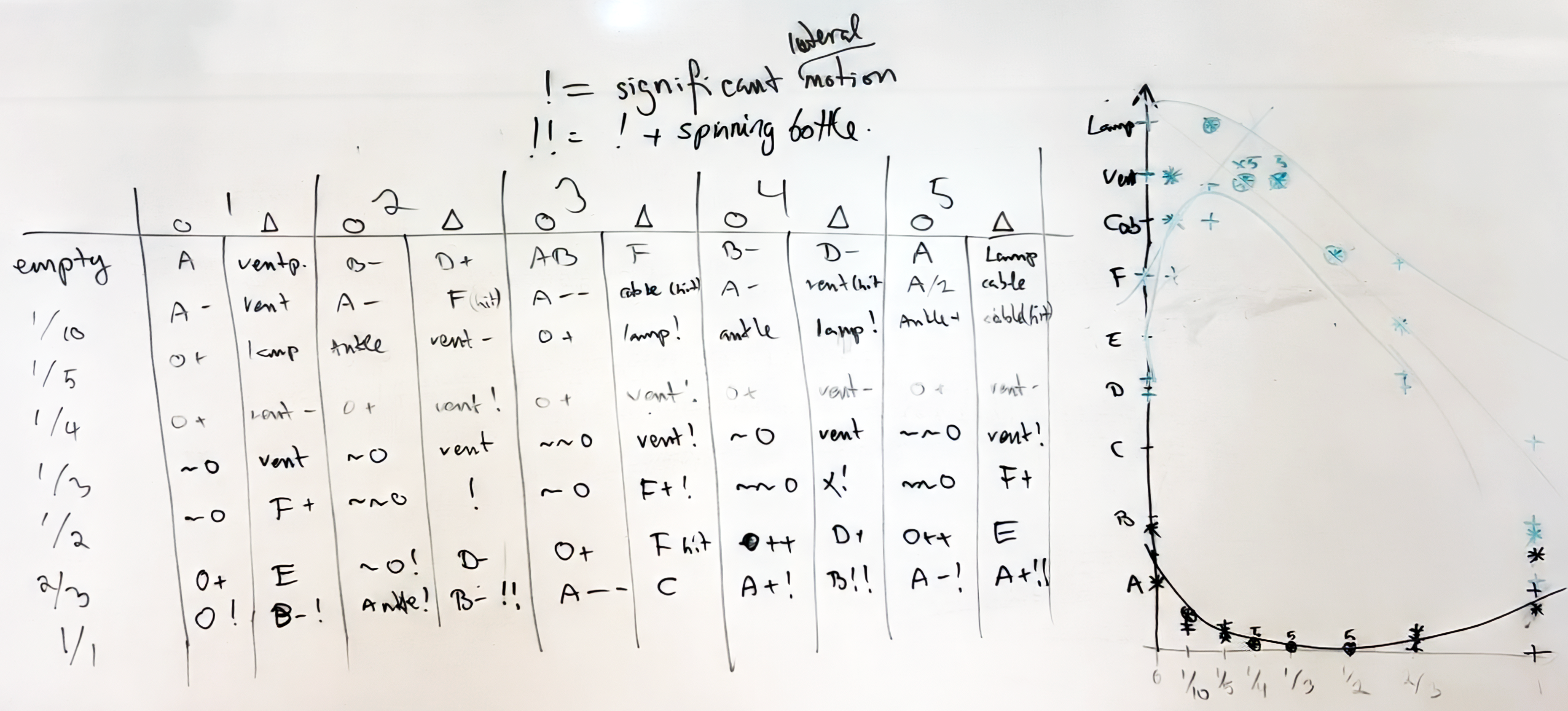}
    \centering
    \caption{The authors' whiteboard resulting from their effort at being students. The resulting plot clearly shows a minimum in the ball bounce height, and that the bottle height data is noisy for low and high masses. Letters correspond to marks on the whiteboards; circle in the column header indicates these are ball heights while a triangle denotes bottle heights.
    \label{fig:OurWhiteboard}
}
\end{figure}

The authors performed two experiments themselves. The first one was as described in section \ref{sec-classroom}. Figure \ref{fig:OurWhiteboard} shows the result from the \textit{qualitative} study, \textit{i.e.}, a deliberate attempt to perform the experiment as if we were students. The vertical scale used was formed of marks on two whiteboards and some fixed items higher up in the room. We also made notes of when the bounce resulted in significant lateral motion. To the right in Fig. \ref{fig:OurWhiteboard} the qualitative results are plotted.

\begin{figure}[htb!]
    \includegraphics[width = \textwidth]{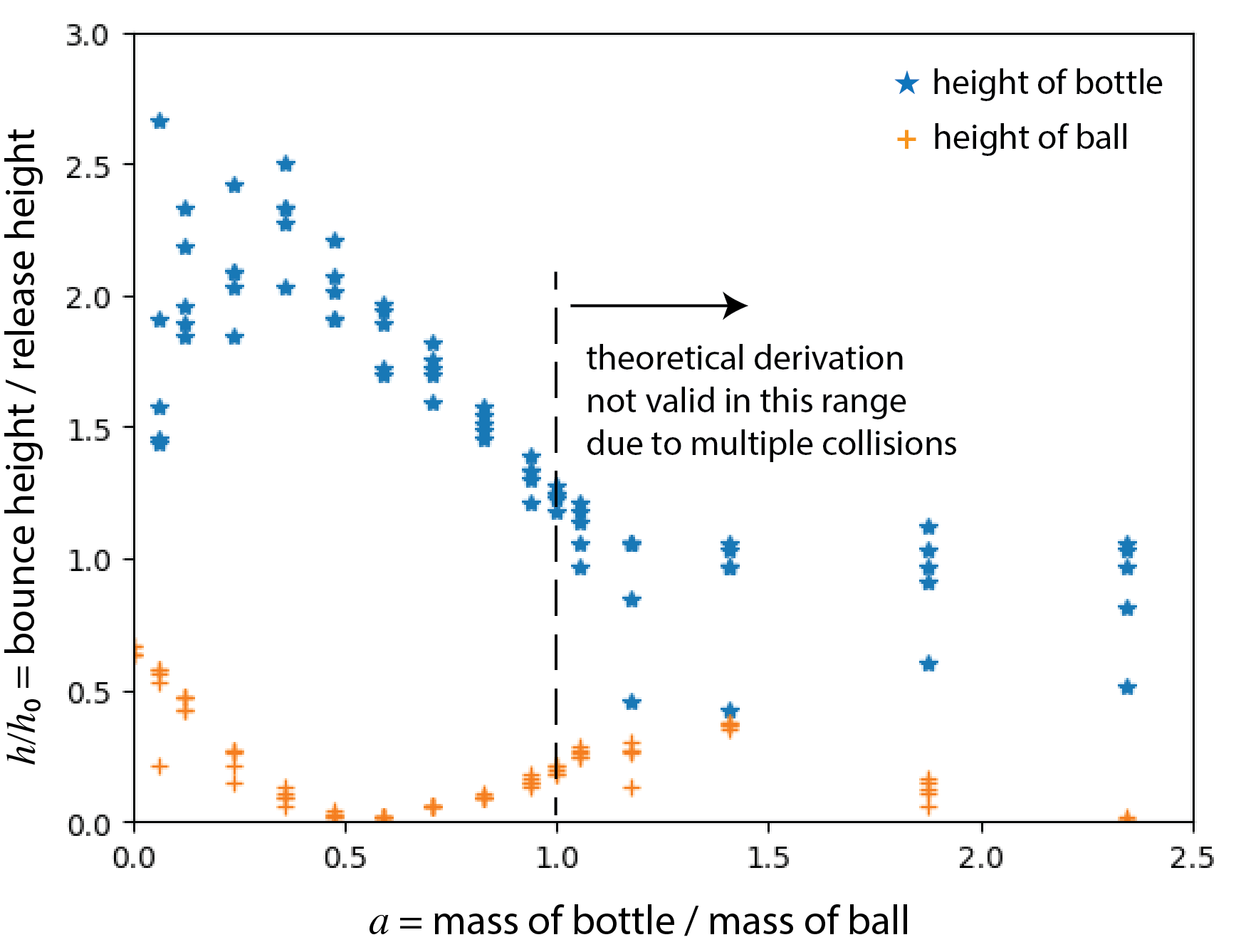}
    \caption{Plot of the result of the quantitative study, showing the bounce height of the ball and bottle as a function of the mass ratio $a=m_{\mathrm{bottle}} / m_{\mathrm{ball}}, \ m_{\mathrm{ball}} = \SI{425}{g} $ and the release height $h_0 \approx \SI{1.4}{m} $. Note that the minimum in the ball bounce occurs at $a \approx 1/2$ and that there is a second maximum at $a \approx 1.5$. The theoretical derivation in \ref{SM-theory} is not valid for $a > 1$ due to multiple collisions. The plot includes the ball bounce with no bottle, yielding $ h/h_0 \approx 2/3 $.
    \label{fig:quant-plot}
    }
\end{figure}

In the second experiment, the aim was to perform a quantitative measurement while still using very simple tools that can be expected to, in principle, be available at most institutions around the world. This quantitative study still did not use a standardised scale (\textit{i.e.} no SI units). We performed the experiment outdoors, using the bricks of a wall as the vertical scale, and video to record the bounces. Instead of water we used air rifle pellets, which allowed us to reach $a \approx 2$ with a \SI{500}{ml} bottle. The release height was $h_0 \approx \SI{1.4}{m}$. The ball and bottle were weighed separately prior to release and the bounce heights subsequently extracted by pausing the digital video and counting bricks; all heights were measured from the underside of the corresponding object. Extracting quantitative data from $5 \times 15 = 75$ runs took around half a day, which is likely to be unfeasibly slow within a class setting, but may be suitable for a project-style implementation where quantifying (as here) follows an initial qualitative effort. Our results are plotted in Fig. \ref{fig:quant-plot}.

Performing a quantitative experiment leads to exactly the same behaviour as seen in the qualitative experiment, but returns the eventual focus to precise measurements, as well as requiring time to extract the data from the videos. Thus, beyond providing a scale for $a$ (which could easily be added to the qualitative case), it adds little, if anything, to the understanding of the phenomenon. In both experiments, a large spread of bottle bounce heights occurs for both low (empty and nearly empty) and high (full) bottle masses. This corresponds to the range where the bottle was difficult to balance on the ball prior to release, requiring adjustments to the method to have a second person balance the bottle. For the higher bottle masses, several collisions and bounces can occur before the objects separate, leading to further variability in the data. For a given value of $a$, lower observed bottle bounce heights involve significant lateral motion (energy and momentum are transferred to horizontal as well as vertical motion).

A more important characteristic shared by the two experiments is that the ball bounce minimum occurs at \textit{higher} mass ratio than the $a = 1/3$ predicted by the theory described above; for $e < 1$ the minimum should actually move to \textit{lower} ratios, $a < 1/3$. We attribute this discrepancy to the fact that the collision in reality is quite far from instantaneous, since the bottle presses down into the ball before separation. This case is not straightforward to treat analytically and requires numerical modelling. A preliminary study using finite element method with explicit time integration found a bounce minimum very close to the bottle mass observed in the data ($a \approx 1/2$). However, the focus of the experiment is on observation and qualitative description of the behaviour -- trusting one's measurements -- and therefore the numerical modelling results are not included in this paper.

\subsection{Student whiteboards}

\begin{figure}[htb!] 
    \includegraphics[width = .6\textwidth, angle = 270, origin=c]{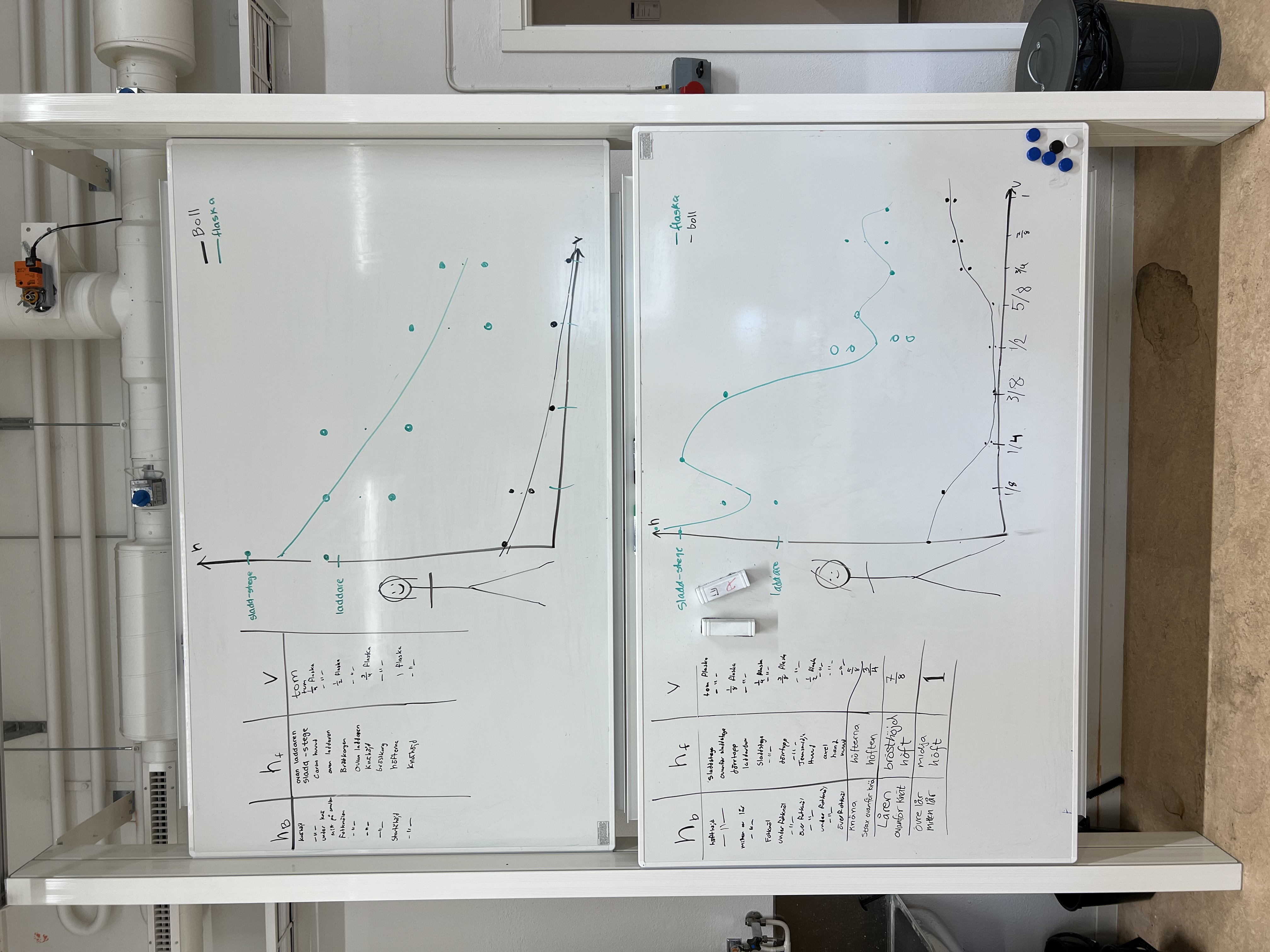}
    \includegraphics[width = .6\textwidth, angle = 270, origin=c]{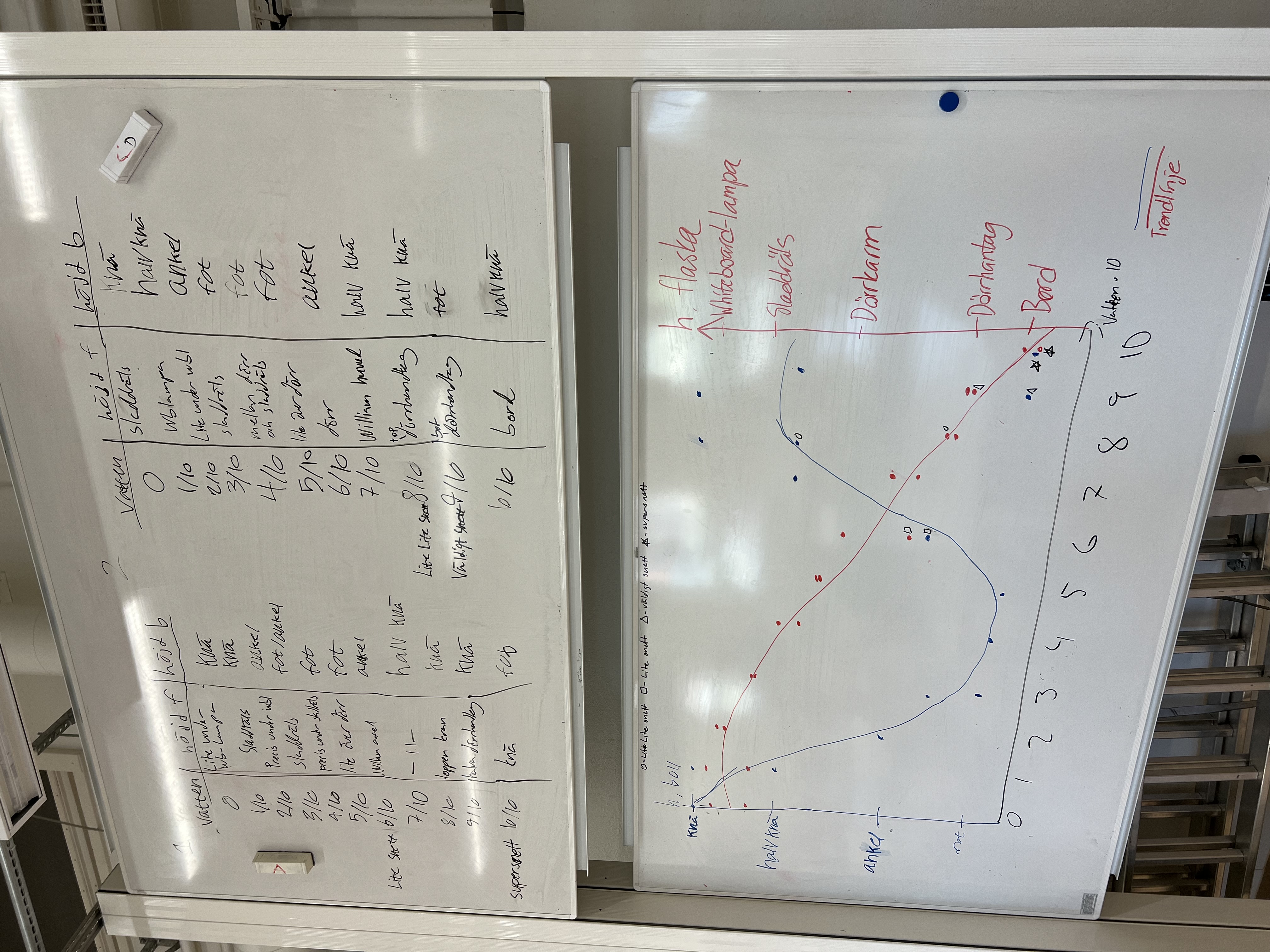}
    \centering
    \caption{
    Two examples of student whiteboards (in Swedish) after the experiment. \textit{Left:} Fairly typical example where the students use a human body as the vertical scale. We also see that they repeated the experiment after being told that there was something more to discover about the ball. \textit{Right:} Here the students, unprompted, used different vertical scales for the bottle and the ball. They also noted some cases where there was significant lateral motion of the bottle (`\textit{snett}' translates to `at an angle').
    \label{fig:studentWB}
    }
\end{figure}

As described above (section \ref{sec-classroom}), the students' task involved scale development, data recording and plotting on whiteboards. Observations could be recorded anywhere -- either on paper or on whiteboards -- and groups were fairly evenly divided between where they chose to record observations (whiteboards, paper, or both). In Fig. \ref{fig:studentWB} we show two good, but characteristic, examples of student whiteboards, showing how this was done, and some of the common practices that could be drawn up in discussions (e.g. joining dots, and that a `sketch' in the context of graph drawing in physics has the specific meaning of drawing a smooth freehand line showing the principal features). These demonstrate how realistic this experiment is within the time available: it is rare that a group does not succeed in collecting data showing the minimum of the ball's bounce (unless they do a different experiment, see below). The observational nature of the experiment means that note keeping is authentic and emphasising it is natural; the whiteboards are also visible to all, and points often unstated can be highlighted through a natural, improvement-focussed commentary.

\section{Pedagogical discussion \label{sec:PedPot}}

\subsection{Characterising inquiry}

We can review the proposed, observation-focussed activity through the rubric used for characterising inquiry \cite{JColSciTeach.38.52}. In Table \ref{tab:CharInq}, the stages of an experiment discussed in Ref. \cite{JColSciTeach.38.52} are given, with most of the combined stages split. For each stage, the students' and teachers' prior and developed knowledge and the critical role of student observations separate to measurement are indicated. The problem, results communication and conclusions are, to some extent, teacher-determined, which are characteristic of `confirmation' activities \cite{JColSciTeach.38.52}. However, students have considerable freedom in the intermediate stages, and there is also no theoretical comparison provided. This marks a significant break with typical frameworks that may be used when designing practical activities, and also indicates that characterising inquiry in terms of material provided and removed sequentially from the \emph{end} of an activity is insufficient and may even be counter to developing activities that encourage the develop of key scientific skills and critical thinking through the provision of genuine and meaningful student agency \cite{PRPER.16.010109, PRPER.17.020128}. Moreover, with no theoretical description or model provided, students cannot, by definition, force their data to match this \cite{PRPER.16.010113, PRPER.17.020112}.

\begin{table}[htb!]
\caption{
Aspects used to characterise the level of inquiry of a practical level (following Ref. \protect{\cite{JColSciTeach.38.52}}), and teacher and student inputs into the activity described. 
\label{tab:CharInq}
}
\begin{tabular}{l|p{5cm}|p{5cm}}
Characteristic & Student Perspective & Teacher Perspective \\ \hline
Problem/Question & Task set by teacher & Given; legitimate alternatives exist \\
Background & Everyday phenomenon; basic idea demonstrated at start of session & Observation \\
Theory & No theory provided or used in the activity & Model developed, but not shared until after all students have completed the practical work \\
Design & Basic idea demonstrated; otherwise open; updated in light of observations & Careful and student-like attempts worked through; some nuances known \\
Procedures & Basic idea demonstrated; otherwise open; updated in light of observations & Some nuances and variations known; prepared to hint for some methodological struggles \\
Results analysis & `Plot on whiteboard' instruction; otherwise open and adjustable & Careful and student-like attempts worked through \\
Results communication & `Plot on whiteboard' instruction; otherwise open & Pointers and advice on results communication (plotting) form part of closing \\
Conclusions & Methodological as well as the behaviour observed; experience based & Model results presented at end, but students' learning from the experience is more in focus \\
\end{tabular}
\end{table}

\subsection{Variations on a theme}
\label{sec:variations}

In our single session (75 minutes plus introduction and final discussion), we emphasise \emph{observation} as an important part of experimental work in physics, along with students' control over (\textit{i.e}. ownership of and agency in \cite{PRPER.17.020128, PRPER.16.010109, ArXiv:2007.07427}) the experiment. The discussion therefore focusses on what students did and observed, including, for example, the need to be consistent in (and record) how the ball was held before release. Feedback focusses on skills aspects such as what is meant by `sketching' in a physics context and how to plot for clarity.

The clarity of instructions appears lead to students all varying the amount of water in the bottle, bur we have also known students in earlier iterations to change the height from which the ball was dropped. While this is not the task described, it is an entirely valid and relevant investigation, and students who start doing this (or similar investigations), should be encouraged to continue, and supported through the fact that they have looked into something quite different. (In varying the release height, students are exploring whether $h/h_0$ is independent of $h_0$, an important point if one wishes to compare results from different groups.)

We used leather footballs, but any ball of a similar size should work, although a lighter ball should ideally be accompanied by a smaller and lighter bottle. Quantifying $a$ (by weighing the ball and the bottle) should be possible without shifting the focus from observation on the bounces: the mass of the bottle is static within an experimental run, so measuring this does not detract from observing the dynamic behaviours in the experiment. However, we found it useful to not reveal the actual value of the ratio of the masses until after the experiment; when asked to guess the ratio, most groups guess that the minimum occurs at $a=1$, leading to an interesting discussion (see \ref{SM-classroom}).

As described above, the physics of the experiment appears to be very simple, but is in reality more complicated since the collision is not strictly instantaneous. As a result, developing a solvable analytical model requires some careful thinking of how the situation can conveniently be described -- i.e. as a two stage process (Figure \ref{fig:BB-illustration}). Thus the experiment can provide plentiful opportunities to discuss the physics and model development that requires that process-based reasoning \cite{LearnInstr.4.27, CogInstr.18.1}. The requisite elements (conservation of energy, conservation of momentum, elasticity) are commonly part of early undergraduate and upper secondary school physics curricula, and the experiment has potential to form the basis of investigative project work at low cost. Since an analytical model describes the behaviour, but not the measurements, a project form of the experiment would have very similar potential to the \textit{RC} circuit experiment described in Ref. \cite{EJP.45.055201}. However, modelling would most likely be restricted to obtaining qualitative (e.g. Figure \ref{fig:Diagram-TheoryPlot}) rather than quantitative agreement. In our short, single laboratory session, we believe that a focus on observation and the experimental process provides a richer learning opportunity. The absence of an instructor-provided theoretical description prevents students from referring to values from a model that they may view as more credible than the data from their experiment which should be trusted. 

While we ran this experiment indoors, with 4 groups in a large ($9 \times $\SI{11}{m}), high-ceilinged room, working outdoors provides more space and may simplify safety aspects. The downsides are that note keeping is most easily done with tables and chairs, and moving whiteboards adds logistical complications (for our university students, most groups used two large whiteboards). Working outdoors also introduces high weather-dependence on both comfort and note keeping. However, outdoors more aspects become available for consideration and exploration, for example, whether the surface on to which the ball lands is smooth, flat and hard like a classroom floor, or softer and uneven, such as grass. This can lead to further opportunities for students to perform a systematic investigation without needing to measure (too) exactly; it also has the potential for whole-class investigations with different groups providing data collected under different conditions that can then be compared to provide a more detailed picture and understanding \cite{JGeogHE.47.677}. A further factor that can be changed is the use of different liquids, for example, the use of carbonated water where students may expect that the behaviour is affected by the increased pressure in the bottle (the key point is the mass), or denser, more viscous fluids such as (silicone) oil or treacle, which avoids liquid sloshing.

\section{Conclusions}
Dropping a football with a bottle of water balanced on top of it can be the basis of a simple experiment well-suited to emphasising observations and note keeping. The experiment is straightforward to perform, and data collected without the use of standardised scales is sufficient to reveal interesting behaviour, specifically the presence of a minimum in the ball bounce height. In addition, while a simple analytical model reproduces the qualitative behaviour, the values of the bottle/ball mass ratio are not the same as those observed when conducting the experiment. The experiment lends itself to student independence and innovation since there are more variables beyond the amount of water in the bottle that can be varied (e.g. the release height, ball pressure), although these do not affect the qualitative behaviour. Other variations include the release (e.g. if the ball is balanced on a hand, drawing the hand out from under the ball can introduce rotational motion), or dropping the bottle onto a stationary ball. Since minimal written instructions are required, and students can be expected to encounter many small challenges and make numerous observations, some of which will require that they adjust their method, the experiment is highly amenable to introducing laboratory note keeping in a genuine, non-formulaic way.

\section*{Ethics statement}
All example whiteboards are from students who agreed to photos being taken of their data for pedagogical development work, including potential publication.

\section*{Original data}
Original data from this study will be made available upon request to the corresponding author. This includes the videos from the quantitative study but not unpublished photos of student whiteboards.

\section*{Acknowledgements}
We thank Dr. Tommy Holmquist for experimental assistance and valuable discussions, Dr. Paul Håkansson for creating and running a 3D FEM model, and Mr. Jens Magnusson for drawing attention to the phenomenon and getting the ball rolling.

\section*{References}

\bibliographystyle{unsrt}

\begin{thebibliography}{10}

\bibitem{SciEd.88.28}
A.~Hofstein and V.~N. Lunetta.
\newblock The laboratory in science education: Foundations for the twenty-first
  century.
\newblock {\em Sci. Educ.}, 88(1):28--54, 2004.

\bibitem{RevEdRes.52.201}
A.~Hofstein and V.~N. Lunetta.
\newblock The role of the laboratory in science teaching: Neglected aspects of
  research.
\newblock {\em Rev. Educ. Res.}, 52(2):201--217, 1982.

\bibitem{ChemEdResPract.8.212}
M.~E. Fay, N.~P. Grove, M.~H. Towns, and S.~L. Bretz.
\newblock A rubric to characterize inquiry in the undergraduate chemistry
  laboratory.
\newblock {\em Chem. Educ. Res. Pract.}, 8:212--219, 2007.

\bibitem{JColSciTeach.38.52}
L.~B. Buck, S.~L. Bretz, and M.~H. Towns.
\newblock Characterizing the level of inquiry in the undergraduate laboratory.
\newblock {\em J. Coll. Sci. Teach.}, 38(1):52--58, 2008.

\bibitem{JEngEd.100.540}
M.~Koretsky, C.~Kelly, and E.~Gummer.
\newblock Student perceptions of learning in the laboratory: Comparison of
  industrially situated virtual laboratories to capstone physical laboratories.
\newblock {\em J. Eng. Educ.}, 100(3):540--573, 2011.

\bibitem{PRPER.18.010128}
C.~Walsh, H.~J. Lewandowski, and N.~G. Holmes.
\newblock Skills-focused lab instruction improves critical thinking skills and
  experimentation views for all students.
\newblock {\em Phys. Rev. Phys. Educ. Res.}, 18:010128, 2022.

\bibitem{PRPER.16.010113}
E.~M. Smith, M.~M. Stein, and N.~G. Holmes.
\newblock How expectations of confirmation influence students' experimentation
  decisions in introductory labs.
\newblock {\em Phys. Rev. Phys. Educ. Res.}, 16:010113, 2020.

\bibitem{PNAS.112.11199}
N.~G. Holmes, C.~E. Wieman, and D.~A. Bonn.
\newblock Teaching critical thinking.
\newblock {\em Proc. Nat. Acad. Sci.}, 112(36):11199--11204, 2015.

\bibitem{PRPER.17.020112}
A.~M. Phillips, M.~Sundstrom, D.~G. Wu, and N.~G. Holmes.
\newblock Not engaging with problems in the lab: Students' navigation of
  conflicting data and models.
\newblock {\em Phys. Rev. Phys. Educ. Res.}, 17:020112, 2021.

\bibitem{PRPER.12.020132}
B.~R. Wilcox and H.~J. Lewandowski.
\newblock Open-ended versus guided laboratory activities: {Impact} on students'
  beliefs about experimental physics.
\newblock {\em Phys. Rev. Phys. Educ. Res.}, 12:020132, 2016.

\bibitem{PRX.10.011029}
E.~M. Smith, M.~M. Stein, C.~Walsh, and N.~G. Holmes.
\newblock Direct measurement of the impact of teaching experimentation in
  physics labs.
\newblock {\em Phys. Rev. X}, 10:011029, 2020.

\bibitem{IntJScholTandL.3.2.16}
C.~Gormally, P.~Brickman, B.~Hallar, and N.~Armstrong.
\newblock Effects of inquiry-based learning on students' science literacy
  skills and confidence.
\newblock {\em Int. J. Schol. Teach. Learn.}, 3(2):16, 2009.

\bibitem{PRPER.16.010109}
N.~G. Holmes, B.~Keep, and C.~E. Wieman.
\newblock Developing scientific decision making by structuring and supporting
  student agency.
\newblock {\em Phys. Rev. Phys. Educ. Res.}, 16:010109, 2020.

\bibitem{PRPER.17.020128}
Z.~Y. Kalender, E.~Stump, K.~Hubenig, and N.~G. Holmes.
\newblock Restructuring physics labs to cultivate sense of student agency.
\newblock {\em Phys. Rev. Phys. Educ. Res.}, 17:020128, 2021.

\bibitem{PRPER.13.010129}
N.~G. Holmes, J.~Olsen, J.~L. Thomas, and C.~E. Wieman.
\newblock Value added or misattributed? {A} multi-institution study on the
  educational benefit of labs for reinforcing physics content.
\newblock {\em Phys. Rev. Phys. Educ. Res.}, 13:010129, 2017.

\bibitem{PRPER.16.020162}
N.~G. Holmes and H.~J. Lewandowski.
\newblock Investigating the landscape of physics laboratory instruction across
  {North America}.
\newblock {\em Phys. Rev. Phys. Educ. Res.}, 16:020162, 2020.

\bibitem{PRPER.18.020129}
A.~Werth, J.~R. Hoehn, K.~Oliver, M.~F.~J. Fox, and H.~J. Lewandowski.
\newblock Instructor perspectives on the emergency transition to remote
  instruction of physics labs.
\newblock {\em Phys. Rev. Phys. Educ. Res.}, 18:020129, Nov 2022.

\bibitem{EJP.40.015702}
K.~Dunnett, M.~N. Gorman, and P.~A. Bartlett.
\newblock Assessing first-year undergraduate physics students' laboratory
  practices: seeking to encourage research behaviours.
\newblock {\em Eur. J. Phys.}, 40(1):015702, 2018.

\bibitem{TPT.53.352}
N.~G. Holmes and D.~A. Bonn.
\newblock {Quantitative Comparisons to Promote Inquiry in the Introductory
  Physics Lab}.
\newblock {\em The Physics Teacher}, 53(6):352--355, 09 2015.

\bibitem{ProcPERC.2020.539}
M.~Sundstrom, A.~M. Phillips, and N.~G. Holmes.
\newblock Problematizing in inquiry-based labs: How students respond to
  unexpected results.
\newblock In Steven Wolf, Michael Bennett, and Brian Frank, editors, {\em
  Physics Education Research Conference, PERC 2020}, Physics Education Research
  Conference Proceedings, pages 539--544. American Association of Physics
  Teachers (AAPT), 2020.

\bibitem{EJP.45.055201}
F.~V. Kowalski.
\newblock The process of constructing new knowledge: an undergraduate
  laboratory exercise facilitated by a vacuum capacitor-resistor circuit.
\newblock {\em Eur. J. Phys.}, 45(5):055201, 2024.

\bibitem{DeFeo2021}
D.~J. DeFeo, T.~C. Tran, and S.~Gerken.
\newblock Mediating students' fixation with grades in an inquiry-based
  undergraduate biology course.
\newblock {\em Sci. \& Educ.}, 30(1):81--102, 2020.

\bibitem{EdPsy.41.75}
P.~A. Kirschner, J.~Sweller, and R.~E. Clark.
\newblock Why minimal guidance during instruction does not work: An analysis of
  the failure of constructivist, discovery, problem-based, experiential, and
  inquiry-based teaching.
\newblock {\em Educ. Psy.}, 41(2):75--86, 2006.

\bibitem{EdPsych.42.99}
C.~E. Hmelo-Silver, R.~Golan~Duncan, and C.~A. Chinn.
\newblock {Scaffolding and Achievement in Problem-Based and Inquiry Learning: A
  Response to Kirschner, Sweller, and Clark}.
\newblock {\em Educ. Psy.}, 42:99--107, 2007.

\bibitem{PRPER.17.010131}
F.~La~Braca and C.~S. Kalman.
\newblock Comparison of labatorials and traditional labs: The impacts of
  instructional scaffolding on the student experience and conceptual
  understanding.
\newblock {\em Phys. Rev. Phys. Educ. Res.}, 17:010131, 2021.

\bibitem{TPT.57.296}
N.~G. Holmes and E.~M. Smith.
\newblock {Operationalizing the AAPT Learning Goals for the Lab}.
\newblock {\em Phys. Teach.}, 57(5):296--299, 2019.

\bibitem{Biggs-and-Tang}
J.~B. Biggs and C.~Tang.
\newblock {\em Teaching for quality learning at university: What the student
  does}.
\newblock McGraw-Hill Education, UK, 4th edition, 2011.

\bibitem{AffectRes.2022.35}
E.~E. Neenan.
\newblock Writing and structure.
\newblock In A.~G. Gibson, editor, {\em The Affective Researcher (Great Debates
  in Higher Education)}, pages 35--62. Emerald Publishing Limited, Leeds, 2022.

\bibitem{PRPER.12.020129}
J.~T. Stanley and H.~J. Lewandowski.
\newblock Lab notebooks as scientific communication: Investigating development
  from undergraduate courses to graduate research.
\newblock {\em Phys. Rev. Phys. Educ. Res.}, 12:020129, 2016.

\bibitem{AlfonsMagnus-2023}
{Andersson, M. and Nordlund, A.}
\newblock {Det kommer in någon form av magkänsla, och det låter ju inte
  riktigt klokt [One has some kind of gut feeling, and it makes absolutely no
  sense]}, {2023}.
\newblock {Student Paper}.

\bibitem{WalkerResnick-book}
J.~Walker and R.~Resnick.
\newblock {\em Fundamentals of physics}.
\newblock Wiley, Hoboken, NJ, 8th edition, 2008.

\bibitem{ArXiv:2007.07427}
Z.~Y. Kalender, M.~Stein, and N.~G. Holmes.
\newblock Sense of agency, gender, and students' perception in open-ended
  physics labs.
\newblock {\em arXiv}, 2020.
\newblock arXiv:2007.07427.

\bibitem{LearnInstr.4.27}
M.~T.~H. Chi, J.~D. Slotta, and N.~{De Leeuw}.
\newblock From things to processes: A theory of conceptual change for learning
  science concepts.
\newblock {\em Learn. Instr.}, 4(1):27--43, 1994.

\bibitem{CogInstr.18.1}
M.~Reiner, J.~D. Slotta, M.~T.~H. Chi, and L.~B. Resnick.
\newblock Naive physics reasoning: A commitment to substance-based conceptions.
\newblock {\em Cognition and Instruction}, 18(1):1--34, 2000.

\bibitem{JGeogHE.47.677}
S.~Praskievicz.
\newblock Field-based local stream research in undergraduate classes: an
  inquiry-based approach.
\newblock {\em Journal of Geography in Higher Education}, 47(4):677--684, 2023.

\end{thebibliography}

\newpage

\setcounter{equation}{0}
\renewcommand{\theequation}{A\arabic{equation}}
\renewcommand{\thesubsection}{A.\Roman{subsection}}

\section*{Appendix}

\subsection{The printed instruction given to students} \label{SM-student-info}

This very simple experiment still contains a lot of physics. In this part, you are to make notes on paper as well as on the whiteboard.

The principle of the experiment is based on the following observation: If you place an empty PET bottle on top of a soccer ball and drop both simultaneously, the bottle will fly upwards much higher than the ball bounces. You will now study this phenomenon \textit{without} making numerical measurements.

\textbf{Aims of the experiment:}
\begin{itemize}
    \item Experimental process – strategy, observation, adjustment, new observation…
    \item Structured notes – what should be noted down? There are headings in the notebook template to guide you
    \item Safety – it’s all about being sensible, pausing, and thinking before you start
\end{itemize}

\textbf{Workflow}
\begin{outline}
    \1 First and foremost: think through the safety aspects – a PET bottle can fly quite far and fast, and we don’t want you to injure yourselves, other people, or equipment.
    \1 Take notes throughout the entire experiment
        \2 The important thing is that you write things down, not exactly how you do it
        \2 What you don’t write down, you won’t remember!
    \1 Design and carry out an experiment to study how high both the ball and the bottle bounce as a function of how much water is in the bottle
        \2 You do not have access to measuring instruments – `measure' using estimation by eye, approximations, marks on a whiteboard, or comparisons
        \2 Invent an arbitrary scale, which you may need to adjust
        \2 Also note if the bottle or the ball bounces at an angle or if something else strange happens
        \2 Plot your `data' on the whiteboard, take a picture of it when you are done
\end{outline}

\subsection{Teachers' `guide'} \label{SM-classroom}

As described in the main text, the students are divided into groups and given the following instructions (also found in the printed lab manual):

\bit
    \item Consider the safety aspects of the experiment. The bottle may bounce unpredictably. Secure laptops and other fragile items.
    \item Document a simple experimental plan in a notebook; be prepared to modify the plan as needed.
    \item Perform the experiment with different amounts of water in the bottle.
    \item Observe both the bottle and the ball; try to discover something interesting.
    \item  \label{point:qual-defs} This is a qualitative experiment. Precise measurements are not allowed; use descriptive terms such as `lamp', `table', `knee', or `ankle' for your data points.
    \item Record your data and observations on a whiteboard, and make notes of any changes to your plan based on what you observe.
    \item Create a visual representation by plotting your qualitative data on the whiteboard.
\eit

\noindent While students are performing the experiment, these are things we commonly prompt them on:
\bit
    \item Take notes, not only data points but also of other observations.
    \item Record changes to methods.
    \item Sketches e.g. of distances.
    \item Releasing/balancing the bottle works better when there are two people -- ask: `might this be easier if you had an extra hand?'
    \item If they seem to not notice the ball bounce minimum, repeat the experiment, possibly for more bottle masses.
    \item But do not guide them too much -- it is fine if a few groups get a bit lost so long as they make progress.
\eit

\noindent After the experiment, all groups in the session are gathered for a closing discussion. The key points we typically cover in the discussion are:

\bit
    \item What did you learn?
    \item Discuss good practices in plotting and graph sketching (see \ref{sec:variations}).
    \item Engage in a qualitative discussion about the underlying physics, essentially conveying the essence of the theoretical derivation \ref{SM-theory}, including a demonstration of the minimum ball bounce.
    \item Allow students a minute to discuss to guess the mass ratio between the ball and the bottle at the point of minimal bounce (most guess 1:1).
    \item Reveal that a full bottle is roughly the same mass as a football. This also provides an opportunity to discuss why it is difficult to measure things using only our body (hands are sensitive mainly to pressure, not weight).
    \item Discuss the expected outcome when the bottle is dropped onto a ball that is stationary on the ground. Demonstrate the result (a full bottle now produces no bounce, whereas a bottle one-third full results in a noticeable bounce).
    \item Ask the students (discuss in groups) what they thought about the experiment. (Most respond that they enjoyed it, noting that they appreciated the freedom to design the experiment, observe, and discover.)
\eit

\subsection{Theoretical derivation} \label{SM-theory}

The following derivation assumes that all collisions are central and instantaneous, and we treat the objects as point masses. The ball and bottle are released together from height $h_0$, and we assume that there is a small separation between them. We set the ball's mass to $m_\mathrm{ball} = m $ and the bottle's mass to $m_\mathrm{bottle} = a m$ where $a > 0$ ($a$ can also be greater than 1). The ball and the bottle both have the speed $-v_0$ when the ball hits the ground. Fig. \ref{fig:BB-illustration} in the main text shows how the experiment is set up and defines the quantities involved. Upward speeds are defined as positive.

After the ball bounces on the ground, but before the collision between bottle and ball, the ball has a speed of $ev_0$ upwards (losing speed in the instantaneous bounce on the ground, $0<e<1$) while the bottle continues to fall with $-v_0$ downwards. The subsequent collision between the ball and the bottle is also instantaneous and inelastic with a coefficient of restitution $\eta$. After the collision, the bottle's velocity is $v_1$ and the ball's is $v_2$ (recall that positive velocity means upwards).

\noindent Conservation of momentum gives:

\beq   
    -amv_0+mev_0=amv_1 + mv_2 \implies v_0(e-a)=av_1 + v_2
    \label{eq-momentum}
\eeq

\noindent The coefficient of restitution is defined as $ |\Delta v_{after}| = \eta \cdot |\Delta v_{before}| $, thus:

\beq
    |v_1-v_2|=\eta \cdot |-v_0-ev_0|
    \label{eq-elastic}
\eeq

\noindent For small $a$, $v_1 > v_2$, and we can thus combine equations \ref{eq-momentum} and \ref{eq-elastic} to get:

\[
\begin{array}{rcl}
    v_1 & = & v_2+ \eta (1+e)v_0 \\
    v_0(e-a) & = & a\left(v_2+\eta(1+e)v_0\right)+v_2 \\
              & = & (1+a)v_2 + a\eta(1+e)v_0 \\
    v_2 & = & \dfrac{e - a - a \eta - a e \eta}{1 + a} v_0 \\
    v_1 & = & \dfrac{e - a - a \eta - a e \eta}{1 + a} v_0 + \eta (1 + e) v_0 \\
         & = & \dfrac{e - a - a \eta - a e \eta + a \eta + \eta + a e \eta + e \eta}{1 + a} v_0 \\
         & = & \dfrac{e + e \eta + \eta - a}{1 + a} v_0 \\
\end{array}
\]

\noindent Finally:

\beq
    v_1 = \frac{e \eta+\eta+e-a}{1 + a} v_0 \ , \quad v_2 = \frac{e-a(1 + e \eta +  \eta)}{1 + a} v_0
\eeq

\subsection*{Observations}
\bit
    \item In the limit $ a = 0 $, the bottle's velocity becomes $ v_1=(ec+e+\eta) v_0 $ or $ 3v_0$ for $e=\eta=1$, \textit{i.e.} three times the speed of the ball (which retains $v_2=ev_0$).
    \item The ball's velocity becomes zero $(v_2=0)$ when the bottle's mass is
    \begin{equation*}
        am = \frac{e}{1+\eta+e \eta} m \implies a = \frac{1}{3} \ \mathrm{for} \ e=\eta=1
    \end{equation*}
    \item For $a > \dfrac{e}{1+\eta+e \eta}$ the ball's velocity becomes negative meaning that it bounces on the ground once again. In the following, we treat this extra bounce as perfectly elastic for simplicity.
    \item In the cases where $|v_2| > |v_1|$, or $a>\dfrac{2e + e \eta+\eta}{2 + e \eta+\eta}$ ($a > 1$ for $e=\eta=1$), the derivation above still holds in principle, since $v_1 > -|v_2|$. However, multiple collisions between the ball and bottle will occur, which is difficult to calculate. In practice this will lead to unpredictable results since the ball and bottle are likely to not be perfectly aligned vertically.
\eit

\subsection*{Bounce height}

The derivation above is for the bounce \textit{speed}. To get the bounce \textit{height}, which is what we see and measure, we use $v=gt$ and $ h=gt^2 / 2 $ which gives $h=v^2 / 2g$.

The speed before collision is $ v_0=\sqrt{2 h_0 g} $, where $h_0$ is the original drop height. And thus we get the bottle and ball bounce heights as:

\beq
    h_1= \left( \frac{e \eta+\eta+e-a}{1 + a} \right)^2 h_0, \quad h_2 = \left( \frac{e-a(1+e \eta+ \eta)}{1 + a} \right)^2 h_0
    \label{full-solution}
\eeq

\subsubsection*{We note that for $e=\eta=1$:}
\bit
    \item The bottle bounce height becomes
    \begin{equation*}        
        h_1= \left( \frac{3-a}{1 + a} \right)^2 h_0 \implies h_1= 9 h_0 \ \mathrm{for} \ a=0
    \end{equation*}
    An empty bottle thus bounces up to nine times higher than the ball!
    \item The ball bounce height becomes
    \begin{equation*}        
        h_2 = \left( \frac{1-3a}{1 + a} \right)^2 h_0 \implies h_2=0 \ \mathrm{for} \ a=\frac{1}{3}
    \end{equation*}
\eit

To measure $e$ we bounce the ball on its own, with $e=\sqrt{h/h_0}$. It is not so simple to measure $\eta$, so as an assumption we use $\eta=e$. In Fig. \ref{fig:Diagram-TheoryPlot} in the main text, solution \ref{full-solution} is plotted with $h_0=1$ and for $e=\eta= \{ 1 \ , \ \sqrt{2/3} \} $. This value of $e$ is based on the quantitative experiment, cf. Fig. \ref{fig:quant-plot}. For bottle masses larger than the minimum point, the ball bounces on the ground a second time; in the derivation above and in the plot, we ignore further losses in the second bounce. We also ignore the fact that the ball and bottle cannot pass each other, which is what the crossing curves at $a \approx 1$ would mean in practice.

\end{document}